# Possible Atmospheric-like Neutron Beams at CSNS

Weijun Ni[a,c)], Hantao Jing[a,b,d,*)], Liying Zhang[a,b,e)], Li Ou[c)]
[a]Institute of High Energy Physics, Chinese Academy of Sciences (CAS), Beijing 100049, China
[b]Dongguan Neutron Science Center, Dongguan 523803, China
[c]Guangxi Normal University, Guilin, 541004, China
[d]State Key Laboratory of Intense Pulsed Radiation Simulation and Effect (North west Institute of Nuclear Technology), Xi'an 710024, China
[e]National Synchrotron Radiation Laboratory, University of Science and Technology of China, Anhui 230029, China

**Abstract:** China Spallation Neutron Source (CSNS) will start commissioning in early 2018. The neutron spectra from the tungsten target bombarding by 1.6 GeV proton beam are very wide, namely white neutron spectra. Two white neutron beams schemed by CSNS are simulated by FLUKA. And the beam spectrum and intensity comparisons with other neutron sources in service are made. The fidelity of neutron spectra of these two white-neutron beam lines for chip irradiation experiments are estimated. The CSNS atmospheric-like neutron beams will be the first white neutron beam lines in China and the most intensive ones in a few years suitable to carry out the accelerated test experiments of neutron single event effect in the world.

## 1 Introduction

The cosmic-ray interacts with air atoms in atmosphere and a large number of neutrons are produced. The neutron fraction in atmospheric radiation field is about 94% at sea level [1] which is however a small percent in outer space. At the same time, the albedo neutrons produced by cosmic-ray interacting with atmospheric constituents will also present near space of 20-100 km from the ground. Therefore, the high-integrated electronic devices on aircrafts in atmosphere and low-orbit satellites near space are subject to meet functional damages of neutron single event effects.

In 1956, W. N. Hess et al. firstly systematically studied the atmospheric neutron environment [2]-[3]. With widespread applications of the large-scale high-integrated electronic devices in the past decades, the neutron single particle effects become more and more remarkable for aircrafts and low-orbit satellites. In late 1980s, IBM and Boeing used aircrafts to study SEE effects of the atmospheric neutron field and the experimental results validates the notability of the neutron field induced single event effects [4]. In 1995, Baumann et al. studied the effects of thermal neutron capture by $^{10}$B in boro-phosphosilicate glass [5]. More recent progresses on neutron single particle effects can be found in [6]-[9].

Usually the atmospheric radiation field is composed of neutrons, protons, alphas, muons, electrons, gammas an so on. In Fig. 1, the integrated flux for several types of main particles will be given. One can find that the neutron intensity is larger than other particles below about 30-km altitude. Obviously the studies for neutron SEE are significant for aircraft and low-orbit satellites. The atmospheric neutron field intensity will be affected by many factors such as time, longitude, latitude, altitude, solar activities and so on. Regardless of the solar high active years, the altitude is the most important by comparing with other factors.

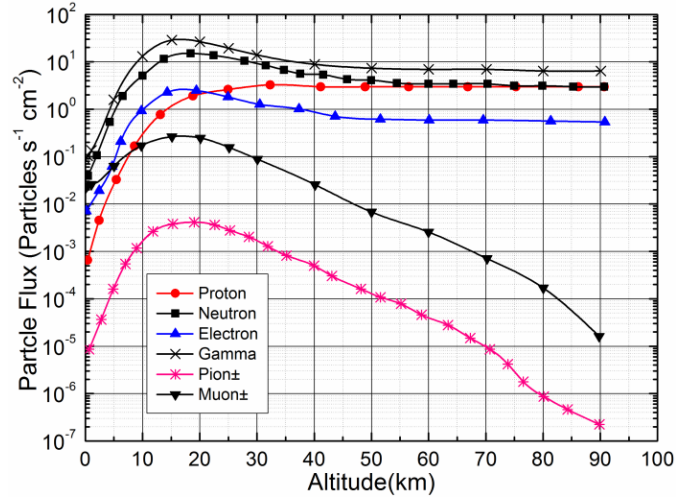

Fig. 1. Particle flux vs. altitude [10].

In atmosphere, the energy spectrum of neutron radiation field is very wide, namely a white neutron spectrum characteristic. Usually the neutrons above 1MeV are remarkable to contribute the SEE. There exist two important reference standard neutron energy spectra for SEE investigations in terrestrial and avionic environments. The JEDEC (formerly known as the Joint Electron Device Engineering Council) provides a reference terrestrial neutron spectrum in JESD89A [11]. The neutron differential flux given in Eq. (1) is based on the reference conditions of sea level in New York City with mid-level solar activity. The total neutron flux of the reference spectrum above 10 MeV is $3.6 \times 10^{-3}$ cm$^{-2}$s$^{-1}$ at NYC [11].

$$\Phi_{JEDEC}(E) = 1.006 \times \exp[-0.35(\ln(E))^2 + 2.1451(\ln(E))] \\ + 1.011 \times 10^{-3} \exp[-0.4106(\ln(E))^2 - 0.667\ln(E)], \quad (1)$$

Where $E$ is the neutron energy and $\Phi_{JEDEC}(E)$ is the reference neutron differential flux. Another authority, namely IEC (the International Electrotechnical Commission) gives another reference neutron flux of IEC TS 62396-1 at 12192 m (40,000 ft), latitude 45° based on 1974 NASA Ames flight data in avionic environments [12]. The parameterized differential flux can be defined as

$$\Phi_{IEC}(E) = 0.346 E^{-0.922} \times \exp[-0.0152(\ln(E))^2] \quad \text{for } E \leq 300 MeV \quad (2)$$
$$= 340 E^{-2.2} \quad \text{for } E > 400 MeV.$$

The real-time tests of electronic devices of course can be obtained the best faithful results for the studies of SEE, but they need a large account of testing time and costs. Therefore, the accelerated tests by atmospheric-like neutron facilities are a good substituted method. At present, there are several important neutron facilities to carry out the accelerated tests of neutron SEE. There are two main different types of facilities to provide high energy neutron beams: 1) spallation neutron source; 2) quasi-mono energetic neutron source. The spallation neutron sources provide neutron beams over a wide range of energies, with the shape of the spectrum similar to that of the terrestrial neutron environment. And a quasi-mono energetic neutron source that may be utilized to calibrate and measure mono-energetic SEU responses at high energies [11].

The Los Alamos Neutron Science Center (LANSCE) is a spallation neutron source by using an 800 MeV proton to bombard tungsten target. The produced neutron energy spectrum covers a range from 1 MeV to 800 MeV [13]. The Svedberg Laboratory (TSL) can supply a quasi-monoenergetic neutron (QMN) beam and a white neutron beam by a 180 MeV spallation source (ANITA) [14]-[16]. The Tri-University Meson Facility (TRIUMF) offers a synthetic neutron field having energies between Thermal to $4 \times 10^8$ eV by using up to a 500-MeV proton beam on a lead or steel target [17]. The Research Center for Nuclear Physics (RCNP) uses a proton beam with incident energies up to 392 MeV on a lead target [18]. The VESUVIO test terminal utilizes one of beam lines from ISIS target station [19]. The high-energy neutrons are few for the VESUVIO because the neutrons are moderated by water. The neutron facility at the Petersburg Nuclear Physics Institute (PNPI) has also developed the neutron beam with the atmospheric-like neutron spectrum [20]-[21]. The characteristic of PNPI neutron spectrum shape is harder and close to 1 GeV at the maximal energy comparing with above neutron sources. The neutron beam energy spectra from these neutron facilities and the standard atmospheric neutron energy spectra of JEDEC and IEC for SEE tests are given in Fig. 2.

China Spallation Neutron Source (CSNS) is a large scientific facility under construction, which is developed mainly for multidisciplinary research on material characterization by using neutron scattering techniques [23]-[25] and expected to be completed in early 2018. It consists of a 1.6-GeV proton accelerator and neutron producing targets. The proton beam bombards a tungsten target with power of 100 kW at phase-I. In the future, the proton beam power will be upgraded to 500 kW. In Fig. 3, the latest CSNS bird's-eye view is given. Besides low-energy neutron beam lines for material studies, two high-energy neutron beam lines has been planned and neutrons are directly extracted from the target rather than from moderators as shown in Fig. 3. One is the back-streaming neutron beam line (Back-n) along the direction of 180-degree about proton beam. It will start commissioning in early 2018. The other is the 41-degree forward neutron beam line which is reserved and can be constructed in the future. In this paper, we will calculate the neutron spectra using the latest FLUKA and evaluate the fidelity of CSNS neutron spectra for SEE accelerated experiments.

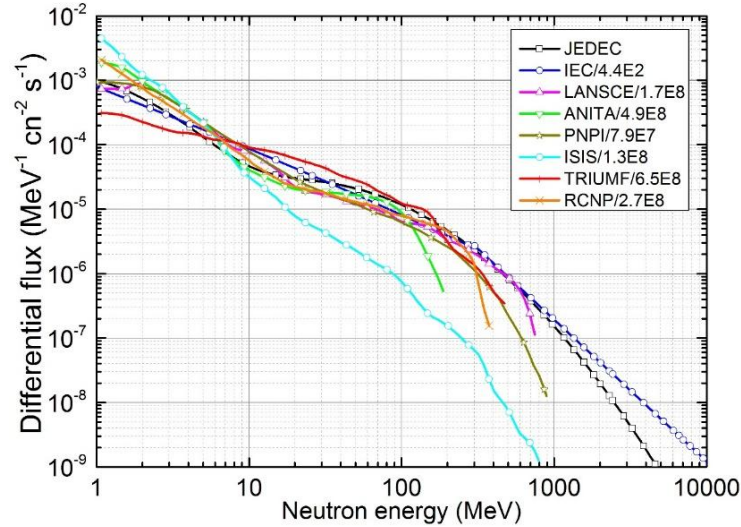

Fig. 2. Standard spectra vs. neutron beam energy spectra from several international atmospheric neutron experimental facilities [22].

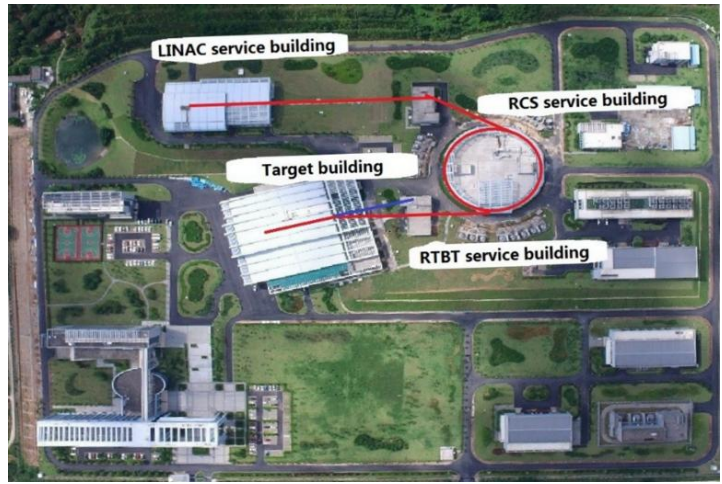

Fig. 3. Latest CSNS bird's-eye view.

## 2  White neutron beams at CSNS

### 1) Spallation target structure and white beam lines at CSNS

The sliced tungsten target (11 pieces, 65 cm in total length, cross-section: $170 \times 70$ mm$^2$) cladded by tantalum (0.3 mm) and cooled by water was employed at CSNS. The target vessel is made of stainless steel (SS316) with a thickness of 2.5 mm for the front side, 7.5 mm for up and down sides and 12 mm for the lateral and back sides. The proton beam spot at the target aiming area is 12 cm (H) $\times$ 4 cm (V) with a quasi-uniform distribution. There are two white neutron beam lines in the proton beam plane: one is the Back-n beam line and the other is a 41° forward beam line (currently sealed and to be opened in the future) designated mainly for high-energy neutron irradiation studies. Between the target-moderator and the biological shielding structure, there are also reflectors of beryllium and iron to enhance the neutron utilization efficiency [26]. Fig. 4 shows the geometry structure of the target region [27]-[29].

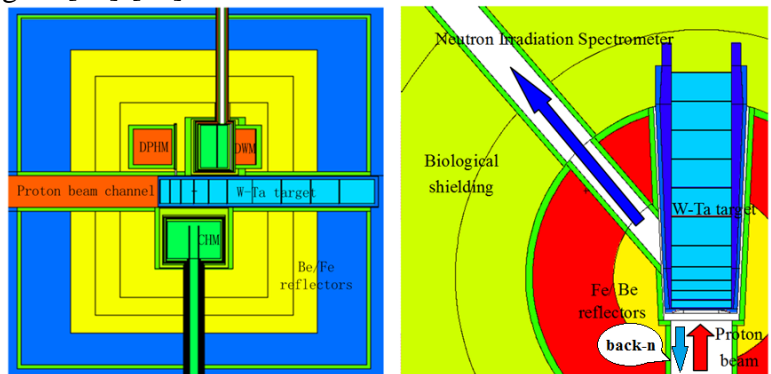

Fig. 4. Geometry of target-moderator-reflector (TMR) model (the left and right are the lateral top views respectively) [27]-[29].

The characteristics of the back-streaming neutrons have been initially studied for the nuclear data measurements in [26]. At present, it is also planned to carry out SEE studies for the Back-n beam line. The layout of the Back-n beam line is presented in Fig. 5. The first 20-m beam line of the Back-n is shared with the last section of the proton beam line – RTBT (Ring to Target Beam Transport). A proton beam window (PBW) in A5083 aluminum alloy of 2 mm in thickness is located at 1.9 m from the target which separates the vacuum tube of the proton beam line and the target vessel. Along the flight path, one neutron shutter (also functioning as a collimator) and two collimators are used to control the neutron beam intensity and beam spot sizes. At about 56 m and 76 m from the spallation target, two endstations (ES#1 & ES#2) will host seven detector systems (or spectrometers) in total for different experiments with only one in a time. ES#1 and ES#2 are used for high-flux experiments and high-resolution experiments, respectively. A preparation room is for preparing experiments or temporal detector storage. An in-room and complex neutron dump is located at rear of ES#2 because of space limitation. In addition, there are three sets of available beam spots, namely Φ30 mm, Φ60 mm and 90 mm×90 mm at ES#2. The neutron intensities corresponding to three sets of beam spots in two endstations are list in Table I. More details about Back-n can be found in [26], [30]-[31]. In order to meet the requirement for the high neutron flux, the SEE accelerated experiments can be arranged in ES#1 (56 m from target) and will share the beam time with other neutron experiments. The scattering foils can be added to zoom out the beam spot size as required. The current schemed fluxes at Back-n can be found in Table I. On the other hand, the length of 41-degree white neutron beam line is about 20 m in Fig. 5. The estimated neutron intensity is about $1.55 \times 10^8$ n/cm$^2$/s at 20 m. This beam line will start to be schemed in the future.

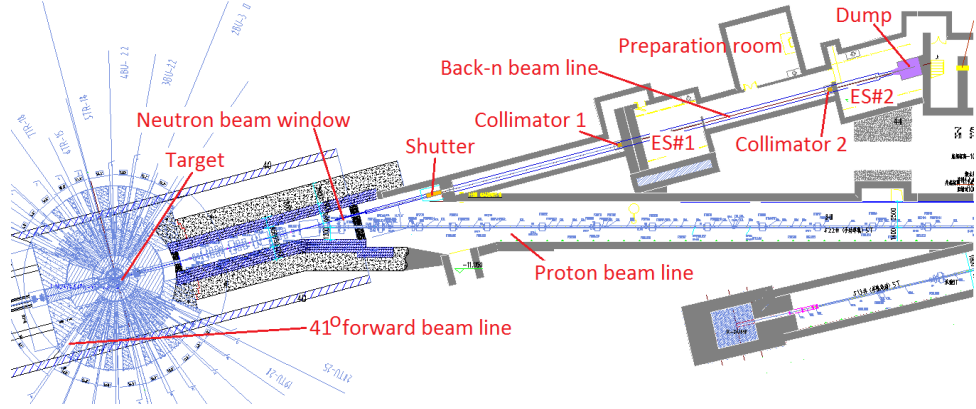

Fig. 5. Layout of the Back-n beam line.

TABLE I
BACK-N NEUTRON BEAM SPOTS & FLUXES AND CORRESPONDING COLLIMATION APERTURE PARAMETERS (100 KW PROTON BEAM) [30].

| ES#2 spot (mm) | Shutter (mm) | Coll#1 (mm) | Coll#2 (mm) | ES#1 spot (mm) | ES#1 flux (/cm$^2$/s) | ES#2 flux (/cm$^2$/s) |
|---|---|---|---|---|---|---|
| ø30 | ø12 | ø15 | ø40 | ø20 | $2.20 \times 10^6$ | $7.81 \times 10^5$ |
| ø60 | ø50 | ø50 | ø58 | ø50 | $4.33 \times 10^7$ | $1.36 \times 10^7$ |
| 90×90 | 78×62 | 76×76 | 90×90 | 75×50 | $5.98 \times 10^7$ | $2.18 \times 10^7$ |

**2) Energy spectrum comparison of CSNS with others**

The simulated white neutron spectra for Back-n and 41º forward beam line of CSNS are given in Fig. 6. At the same time, neutron spectra from other facilities and JEDEC and IEC are compared with CSNS cases. Obviously, all synthetic neutron spectra are somewhat different from JEDEC and IEC standard spectra. Due to characteristics of facilities, the neutron distributions in different energy regions are great different as shown in table II.

In order to evaluate the accelerated-test performances of several neutron facilities, the acceleration factor of A can be defined as [22],

$$A = (\int_{E_{min}}^{\infty} \Phi_{acc}(E) dE) / (\int_{E_{min}}^{\infty} \Phi_{spec}(E) dE), \quad (3)$$

Where $E_{min}$ is the cutoff of a minimal neutron energy required to generate an error event. In our calculations, the $E_{min}$ is assumed as 1 MeV. $\Phi_{acc}(E)$ is the neutron fluxes from atmospheric-like neutron facilities and $\Phi_{spec}(E)$ is the JEDEC or IEC neutron flux. The results are listed in last column in table II. Obviously, the testing acceleration factors for two CSNS beam line are much larger than other likewise testing facilities due to the high-power proton beam and high-yield target at CSNS.

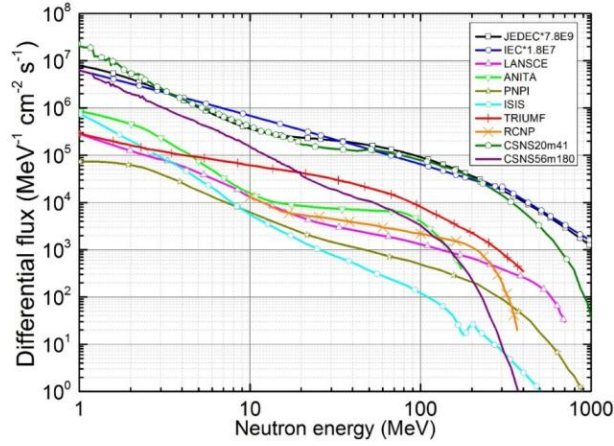

Fig. 6. Energy spectrum comparison of CSNS with others. JEDEC and IEC standard energy spectra respectively time a normalized fractional factor in order to comparing with the CSNS energy spectra.

TABLE II
NEUTRON DISTRIBUTIONS IN DIFFERENT ENERGY REGIONS AND TESTING ACCELERATION FACTORS FOR MAIN FACILITIES IN THE WORLD.

| Source | Neutron fraction and flux | | | | Acceleration factor |
|---|---|---|---|---|---|
| | 1-10 MeV | 10-100 MeV | >100 MeV | Total > 1 MeV (n cm$^{-2}$ hr$^{-1}$) | |
| JEDEC(NYC) | 35% | 35% | 30% | $2.0 \times 10^{1}$ | |
| IEC(12.2 km) | 35% | 35% | 29% | $8.8 \times 10^{3}$ | |
| ANITA | 65% | 28% | 7% | $7.7 \times 10^{9}$ | $4.0 \times 10^{8}$ |
| LANSCE | 52% | 26% | 22% | $3.6 \times 10^{9}$ | $1.8 \times 10^{8}$ |
| TRIUMF | 24% | 54% | 21% | $1.3 \times 10^{10}$ | $6.6 \times 10^{8}$ |
| ISIS | 92% | 7% | 1% | $2.8 \times 10^{9}$ | $1.4 \times 10^{6}$ |
| RCNP | 57% | 25% | 18% | $5.0 \times 10^{9}$ | $2.6 \times 10^{8}$ |
| PNPI | 57% | 29% | 14% | $1.5 \times 10^{9}$ | $7.7 \times 10^{7}$ |
| CSNS 20m @41° | 50% | 29% | 21% | $1.5 \times 10^{11}$ | $7.8 \times 10^{9}$ |
| CSNS 56m @180° | 82% | 17% | 1% | $3.4 \times 10^{10}$ | $1.8 \times 10^{9}$ |

## 3  SER evaluation of CSNS white neutron beams

The soft error is a subset of single event effects. In principal, the soft errors most frequently arise in single event effects. The accelerated tests are usually used to study the soft error of devices under test. The soft error rate (SER) is employed to evaluate the radiation resistance performances. The SER caused by a neutron field can be defined as

$$R = \int_{E_{min}}^{\infty} \sigma(E)\Phi(E)dE, \quad (4)$$

Where $\Phi(E)$ is the differential flux of the neutron field, given in units of neutron number cm$^{-2}$ MeV$^{-1}$ s$^{-1}$. $\sigma(E)$ is the SER cross section with the neutron energy E and given in units of cm$^2$ [22]. Usually the SER cross section can be approximately depicted by a parameterized formula, namely a four-parameter Weibull distribution in JESD89A,

$$\sigma(E) = \sigma_L(1 - e^{-[(E-E_0)/W]^S}), \quad (5)$$

Where $\sigma_L$ is the asymptotic high energy cross section, $E_0$ is the cutoff energy, W is the width parameter, and S is the shape factor. Several typical normalized Weibull approximation of cross-section of $\sigma/\sigma_L$ are given in Fig. 7. Some typical parameters can be found in [22]. The Weibull formulas with two sets of parameters of $E_0$=1 MeV, S=1 and $E_0$=12 MeV, S=3 approximately represent the situations of bit upsets of SRAM and DRAM respectively. And the case with parameters of $E_0$=45 MeV and S=1 approximately represents the situation of single event latchups for both SRAM and DRAM.

Similar to Slayman's methods, one employs the SER ratio to evaluate to the fidelity of energy spectra for a neutron testing beam. It is defined a ratio of the soft error rates from CSNS and other facilities' spectra to that from JEDEC or IEC standard energy spectra, namely

$$\text{SER ratio} = R_{acc}/R_{spec}$$
$$= (\int_{E_{min}}^{\infty} \sigma(E)\Phi_{acc}(E)\, dE) / (A\int_{E_{min}}^{\infty} \sigma(E)\Phi_{spec}(E)\, dE), \quad (6)$$

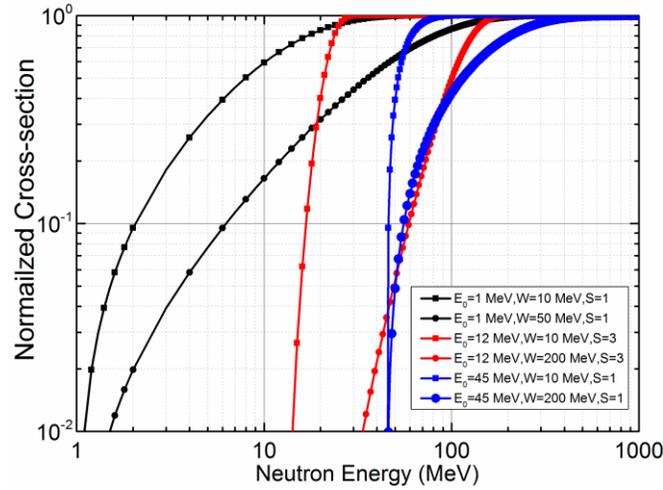

Fig. 7. Examples of normalized Weibull approximation of cross sections for several sets of typically parameters.

Where $\sigma(E)$ are evaluated by parameterized Weibull formula. The real neutron spectra of $\Phi_{acc}(E)$ from different facilities in Fig. 6 is employed. $\Phi_{spec}(E)$ denotes the JEDEC or IEC neutron flux. Normally, SER ratio values greater or less than 1 respectively mean the over-predicting or under-predicting SERs comparing with JEDEC or IEC neutron field.

For above three sets of Weibull parameters, the SER ratios are calculated by using the CSNS neutron fluxes of Back-n and 41 degree beam line. At the same time, the SER ratio results by using neutron flux of other facilities and JEDEC and IEC are also given in Fig. 8 and Fig. 9. The energy spectra of CSNS 41-degree beam line is harder and more faithful. Now the beam line and experimental terminal are being schemed. Although the Back-n spectrum is somewhat softer, it is very similar to ISIS neutron spectra and can also be carry out accelerated tests. The Back-n beam line has been built and will commission in 2018. Furthermore, two CSNS neutron beam lines have the best intensities comparing with other similar facilities in the world.

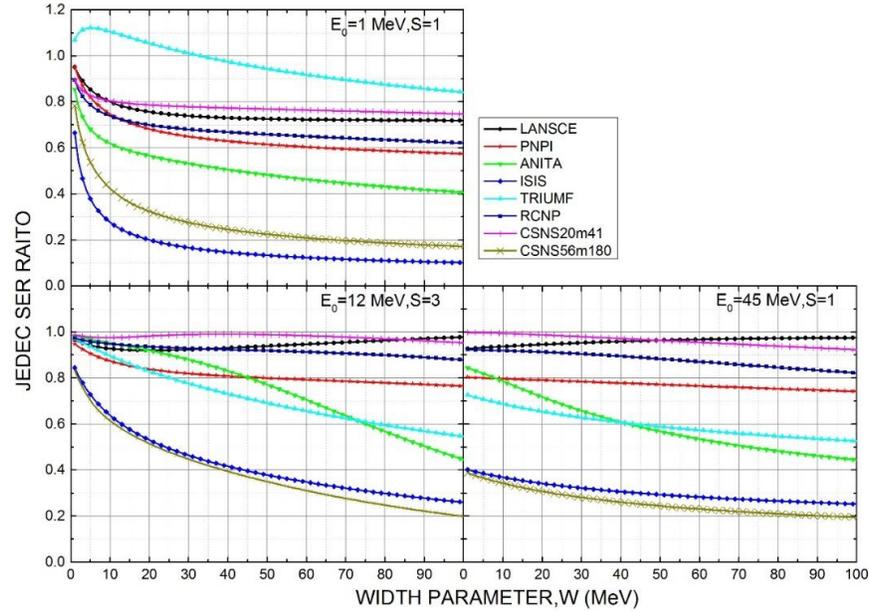

Fig. 8. Comparisons of JEDEC SER ratios using CSNS neutron spectra with respect to others as a function of width parameter for three cases.

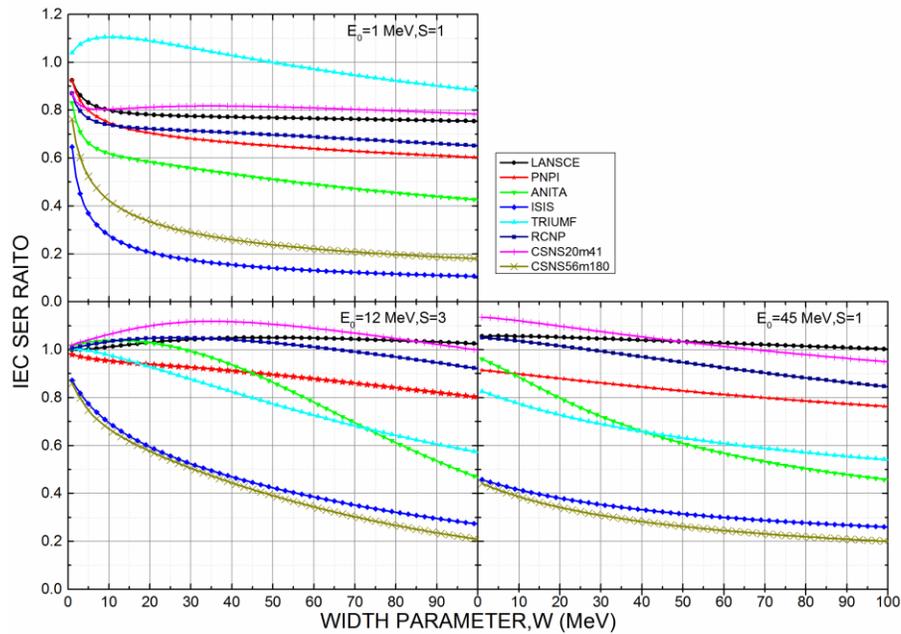

Fig. 9. Same to Fig. 8 but for the IEC neutron spectrum.

## 4    Conclusion

The energy spectrum fidelities of CSNS Back-n and 41-degree forward beam lines for neutron accelerated tests are estimated. At the same time, the comparison of calculated SER by using neutron beam parameters from CSNS and other facilities is made. Obviously, the CSNS white neutron beams are predominant with their high beam intensities. Although the Back-n spectrum shape is softer in the high-energy region, it still can be available to accelerated experiments completely. In the following designs, the experimental conditions and layout in Back-n experimental halls will be planned in details.

In the future, the 41-degree forward beam line will be built in 1-2 years. At present, the CSNS Back-n beam line has been prepared and is debugging. The first neutron at Back-n have been seen in sept. 2017. Although the Back-n is mainly designed for the nuclear data measurements, it is possible to carry out anti-irradiation experiments by sharing the experimental time in 2018.

**Acknowledgements:**

This work was supported by the Opening Special Foundation of State Key Laboratory of Intense Pulsed Radiation Simulation and Effect (Project number SKLIPR.1515) and the National Key Research and Development Program of China (Project: 2016YFA0401600). The authors would like to thank the colleagues of the CSNS target design and white neutron source collaboration for discussions.